\documentclass[
twocolumn,
showpacs,showkeys,preprintnumbers,aps,prd]{revtex4}

\usepackage{amsmath}
\usepackage{amssymb}
\usepackage{revsymb}
\usepackage{graphicx}
\usepackage{dcolumn}

\begin{document}

\title{Axial charges of octet and decuplet baryons}
\author{
Ki-Seok Choi, W. Plessas, and R.\,F. Wagenbrunn}
\affiliation{
Theoretische Physik, Institut f\"ur Physik, Karl-Franzens-Universit\"at,
Universit\"atsplatz 5, A-8010 Graz, Austria}

\begin{abstract}
We present a study of axial charges of baryon ground and resonant states with
relativistic constituent quark models. In particular, the axial charges of octet
and decuplet $N$, $\Sigma$, $\Xi$, $\Delta$, $\Sigma^*$, and $\Xi^*$  baryons
are considered. The theoretical predictions are compared to existing experimental
data and results from other approaches, notably from lattice quantum chromodynamics
and chiral perturbation theory. The relevance of axial charges with regard to
$\pi$-dressing and spontaneous chiral-symmetry breaking is discussed.
\end{abstract}


\maketitle

\section{Introduction}

The axial charges $g_A$ of baryon states are essential quantities for the understanding
of both the electroweak and strong interactions within the Standard Model of
elementary-particle physics. They do not only govern weak processes, such as the
$\beta$ decay, but also intertwine the weak and strong interactions. This is most
clearly reflected by the Goldberger-Treiman relation, which in case of the $N$ reads
$g_A=f_{\pi}g_{\pi NN}/M_N$~\cite{Goldberger:1958tr}. Given the $\pi$ decay constant
$f_{\pi}$ and the nucleon mass $M_N$, the $\pi NN$ coupling constant $g_{\pi NN}$
just turns out to be proportional to $g_A$. Thus the relevance of $\pi$ degrees of
freedom in (low- and intermediate-energy) hadronic physics is intimately tied to the axial
charges: Whenever $g_A$ becomes sizable, $\pi$ degrees of freedom should matter sensibly.
Therefore $g_A$ can also be viewed as an indicator of the phenomenon
of spontaneous breaking of chiral symmetry (SB$\chi$S) of non-perturbative
quantum chromodynamics (QCD),
which is manifested by the non-vanishing value of the light-flavor chiral
condensate $\left<0|q\bar q|0\right>^{1/3}\approx -235$ MeV. The axial charges
thus constitute important parameters for low-energy effective theories. Any
reasonable model of non-perturbative QCD should yield the $g_A$ of correct sizes.
In fact, the axial charges may be considered as benchmark observables
for the nucleon, and more comprehensively the baryon, structures.

Best known is, of course, the axial charge of the $N$, as its experimental value 
can be deduced from the ratio of the axial to the vector coupling
constants $g_A/g_V=1.2695\pm0.0029$~\cite{Amsler:2008zzb}; usually this is done under
the assumption of conserved vector currents (CVC), which implies $g_V=1$. The deviation of
$g_A$ from 1, the axial charge of a point-like particle, can be attributed,
according to the Adler-Weisberger sum rule~\cite{Adler:1965ka,Weisberger:1965hp}, to the
differences between the $\pi^+ N$ and $\pi^- N$ cross sections in pion-nucleon scattering.
Unfortunately, axial charges of other baryon (ground) states are not known directly
from experiment.

The axial charges are also 'measured' in lattice QCD. An increasing number of results
has recently become available, even from full-QCD lattice calculations. A recent summary
of the lattice-QCD results for the $g_A$ of the nucleon is presented in
ref.~\cite{Renner_2009}. The axial charges of hyperons have been studied by Lin et
al.~\cite{Lin:2007ap} as well as Erkol et al.~\cite{Erkol:2009ev} and Engel et
al.~\cite{Engel:2009nh} in (2+1)- and
2-flavor lattice QCD, respectively. There have also been a number of other attempts
to explain the axial charges of the $N$ and the other baryons. We mention only the
more modern ones through chiral perturbation theory
($\chi$PT) (see the recent review by Bernard~\cite{Bernard:2007zu} or, for example,
ref.~\cite{Jiang:2009sf}), chiral unitary
approaches~\cite{Nacher:1999vg}, and relativistic constituent quark models
(RCQM)~\cite{Glozman:2001zc,Boffi:2001zb,Merten:2002nz}. 

Recently, also the axial charges of the $N$ resonances have come into the focus of
interest, because of the question of restoration of chiral symmetry higher in the
baryon (as well as meson) spectra. Specifically, it has been argued that the magnitudes
of $g_A$ should become small for almost degenerate parity-partner $N$ resonances,
indicating the onset of chiral-symmetry restoration with higher excitation
energies~\cite{Glozman:2007ek,Glozman:2008vg}. As the $g_A$ values of $N$
resonances are not known from phenomenology and can hardly be measured experimentally,
this remains a highly theoretical question. However, the problem can be explored 
with the use of lattice QCD. Corresponding first results have already become
available, but only for two of the $N$ resonances, namely, $N$(1535) and
$N$(1650)~\cite{Takahashi:2008fy}. Both of them have spin $J=\frac{1}{2}$ and
parity $P=-1$. Since there is not yet any lattice-QCD result for
positive-parity states, the above issue relating to parity-doubling remains
unresolved from this side. 

The problem of $g_A$ of the $N$ resonances has most recently also been studied
within the RCQM~\cite{Choi2010}. The axial charges of all the $N$ resonances
up to $\sim$1.9 GeV and $J^P=\frac{1}{2}^\pm, \frac{3}{2}^\pm, \frac{5}{2}^\pm$
have been calculated with $N$ resonance wave functions stemming from realistic RCQM
with Goldstone-boson-exchange (GBE) as well as one-gluon-exchange (OGE) hyperfine
interactions. One has found the remarkable result that, especially in case of the
GBE RCQM, the magnitudes of the axial charges need not be small, even if the
energy levels of the opposite-parity partners become (almost) degenerate at
increased excitation energies, e.g. the $J^P=\frac{5}{2}^\pm$ resonances
$N$(1680) and $N$(1675). Thus the issue of possible chiral-restoration phenomena
reflected by the axial charges remains tantalizing until further insights become
available (e.g. from lattice QCD or alternative attempts).

Another question related to the axial charges of the $N$ resonances concerns the
role of $\{QQQQ\bar Q\}$ components. It has been argued that sizable admixtures of
$\{QQQQ\bar Q\}$ are needed in order to reproduce an almost vanishing $g_A$ of the
$N$(1535) resonance~\cite{An:2008tz,Yuan:2009st}. However, these results are usually
obtained in a simplistic non-relativistic approach. Meanwhile it is known that a
RCQM with realistic $\{QQQ\}$ wave functions can easily explain a practically
vanishing $g_A$ of $N$(1535)~\cite{Choi2010}, in perfect congruency with the
predictions obtained from lattice QCD~\cite{Takahashi:2008fy}, and there is no
need for considerable $\{QQQQ\bar Q\}$ admixtures in this case. Moreover, the
correct sizes of the axial charges of the $N$ ground state and the $N$(1535)
as well as $N$(1650) resonances can simultaneously and consistently be reproduced
within a RCQM with only $\{QQQ\}$ configurations~\cite{Choi2010}.

In the context of hyperons the axial charges are also important to learn about the role
of $SU(6)$ flavor-symmetry breaking. In particular, in the case of conserved $SU(3)_F$
the axial charges of the $N$, $\Sigma$, and $\Xi$ ground states are connected by the
following simple relations~\cite{Gaillard:1984ny,Dannbom:1996sh}
\begin{equation}
g_A^N=F+D\,, \hspace{3mm} g_A^{\Sigma}=\sqrt{2}F\,, \hspace{3mm} g_A^{\Xi}=F-D\,,
\label{axcharges}
\end{equation}
which follow through $SU(3)$ Clebsch-Gordan coefficients in the decomposition of the
axial form factor into the functions $F$ and $D$ relating to the octet components in
$SU(3)$~\cite{de Swart:1963gc}. Note that we adopt the convention of $g_A/g_V$ being
positive for the $N$ (like in ref.~\cite{Gaillard:1984ny}) contrary to the
PDG~\cite{Amsler:2008zzb}; this then determines also the signs of all other baryon
axial charges according to Eq.~(\ref{axcharges}).

In the present paper we present results from a comprehensive study of the axial charges
of octet and decuplet ground states $N$, $\Sigma$, $\Xi$, $\Delta$, $\Sigma^*$,
and $\Xi^*$ as well as their resonances along RCQMs. In particular, we employ the RCQMs
whose quark-quark hyperfine interactions derive from OGE~\cite{Theussl:2000sj} and
GBE dynamics~\cite{Glozman:1997fs}; in the latter case we consider both the
version with only the spin-spin interaction from pseudoscalar exchange
(psGBE)~\cite{Glozman:1997ag} as well as the extended version that includes all
force components (i.e. central, tensor, spin-spin, and spin-orbit) from pseudoscalar,
scalar, and vector exchanges (EGBE)~\cite{Glantschnig:2004mu}.
The calculations are performed in the framework of Poincar\`e-invariant quantum mechanics.
In order to keep the numerical computations manageable, we have to restrict the axial 
current operator to the so-called spectator model (SM). It means that the weak-interaction
gauge boson couples only to one of the constituent quarks in the baryon. This approximation
has turned out to be very reasonable already in a previous study of the axial and induced
pseudoscalar form factors of the nucleon~\cite{Glozman:2001zc}, where the SM was employed
specifically in the point form (PF) of relativistic quantum mechanics~\cite{Melde:2004qu}.
It has also been used in studies of the electromagnetic structure of the $N$,
reproducing both the proton and neutron form factors in close agreement with the
experimental data~\cite{Boffi:2001zb,Wagenbrunn:2000es,Berger:2004yi,Melde:2007zz}.

In the following chapter we explain the formalism for the calculation of the matrix
elements of the axial current operator and give the definition of the axial charges
for the different baryon ground and resonant states. Subsequently we
present the results and compare them to experimental data as well as to results from
other approaches, notably from lattice QCD and $\chi PT$. In the final chapter
we draw our conclusions.

\section{Formalism}

In hadronic physics the (diagonal) baryon axial charges $g_A^B$ govern such processes
like $n\rightarrow pe^-\bar \nu_e$, $\Sigma^-\rightarrow \Sigma^0 e^-\bar \nu_e$,
$\Xi^-\rightarrow \Xi^0 e^-\bar \nu_e$ etc. They can generally be calculated through
semileptonic decays $B_1 \rightarrow B_2 \ell\bar \nu$ with strangeness change
$\Delta S=0$. The axial charge is conveniently defined through the value of the axial
form factor $G_A(Q^2)$ at $Q^2=0$, where $Q^2=-q^2$ is the four-momentum transfer.
The axial form factor $G_A(Q^2)$ can be deduced from the relativistically invariant
matrix element of the axial current operator $\hat A^\mu_+ (Q^2)$ sandwiched
between the eigenstates of baryons $B_1$ and $B_2$. Here, the subscript $+$ refers
to the isospin-raising ladder operator
$\tau_+ = \frac{1}{2}\left(\tau_1 + i\tau_2\right)$, with $\tau_i$ being
the usual Pauli matrices. In the specific case of the neutron $\beta$ decay the matrix
element of $\hat A^\mu_+(Q^2=0)$ reads
\begin{equation}
\left<p\left|\hat A^\mu_+\right|n\right>=
g_A^N \bar U_p(P,J'_3)\gamma^\mu \gamma_5 \frac{\tau_+}{2}
U_n(P,J_3) \,,
\label{n-p}
\end{equation}
where $U_n$ and $U_p$ are the neutron and proton spinors, depending on the
four-momentum $P$ and helicities $J_3$ and $J'_3$, respectively; $\gamma^\mu$
and $\gamma_5$ are the usual Dirac matrices. Alternatively, the matrix element in
Eq.~(\ref{n-p}) can also be expressed as
\begin{equation}
\left<p\left|\hat A^\mu_3\right|p\right>=
g_A^N \bar U_p(P,J'_3)\gamma^\mu \gamma_5 \frac{\tau_3}{2}
U_p(P,J_3) 
\label{p-p}
\end{equation}
or
\begin{equation}
\left<n\left|\hat A^\mu_3\right|n\right>=
g_A^N \bar U_n(P,J'_3)\gamma^\mu \gamma_5 \frac{\tau_3}{2}
U_n(P,J_3) \,.
\label{n-n}
\end{equation}

In the spirit of the latter relations we may express the axial charge $g_A^B$ of any
baryon $B=N, \Delta, \Sigma, \Xi, ...$ and its resonances more generally. Let us denote
the baryon states by $\left|B;P,J,J_3\right>$, i.e. as simultaneous
eigenstates of the four-momentum operator $\hat P^\mu$, the intrinsic-spin operator
$\hat J$ and its $z$-projection $\hat J_3$. Since $\hat P^\mu$ and the invariant mass
operator $\hat M$ commute, these eigenstates can be obtained by solving the eigenvalue
equation of $\hat M$
\begin{equation}
\hat M \left|B;P,J,J_3\right>=M \left|B;P,J,J_3\right> \, ,
\end{equation}

Then the axial charge $g_A^B$ of any baryon state $B$ with
$J=\frac{1}{2}, \frac{3}{2}, \frac{5}{2}$ is given by the matrix elements of the axial
current operator $\hat A_3^{\mu}$ for zero momentum transfer $Q^2$ as
\begin{eqnarray}
&&\left<B;P,\frac{1}{2},J_3'\left|{\hat A}^{\mu}_3\right|B;P,\frac{1}{2},J_3\right>=\nonumber\\
&&\hspace{1.1cm}C_B\bar U_B(P,J_3')g_A^B \gamma^{\mu}\gamma_5 \frac{\tau_3}{2} U_B(P,J_3) \, ,\nonumber\\
&&\left<B;P,\frac{3}{2},J_3'\left|{\hat A}^{\mu}_3\right|B;P,\frac{3}{2},J_3\right>=\nonumber\\
&&\hspace{1.1cm}C_B\bar U_B^\nu(P,J_3')g_A^B \gamma^{\mu}\gamma_5 \frac{\tau_3}{2}
U_{B;\nu}(P,J_3) \, ,\nonumber\\
&&\left<B;P,\frac{5}{2},J_3'\left|{\hat A}^{\mu}_3\right|B;P,\frac{5}{2},J_3\right>=\nonumber\\
&&\hspace{1.1cm}C_B\bar U_B^{\nu\lambda}(P,J_3')g_A^B \gamma^{\mu}\gamma_5 \frac{\tau_3}{2}
U_{B;\nu\lambda}(P,J_3) \, , 
\end{eqnarray}
where the coefficients $C_B$ are specified by
\begin{equation}
C_N=2C_{\Delta}=\frac{1}{\sqrt{2}}C_{\Sigma}=C_{\Xi}=1 \,.
\end{equation}
Here, $U_B(P,J_3)$ are the usual Dirac spinors for $J=\frac{1}{2}$ baryons and
$U_{B;\nu}(P,J_3)$ as well as $U_{B;\nu\lambda}(P,J_3)$  are the Rarita-Schwinger
spinors~\cite{Rarita:1941mf} for $J=\frac{3}{2}$ and $J=\frac{5}{2}$ baryons,
respectively, with the normalizations as given in Appendix A.

Omitting from now on the denotation after $B$ we can write the matrix elements of
$\hat A^\mu_3$ for any ground and resonance states as
\begin{eqnarray}
&&\left<P,J,J_3'\right|{\hat A^\mu_3 (Q^2=0)}\left|P,J,J_3\right>=
\nonumber \\
&&
2M\sum_{\sigma_i\sigma'_i}{\int{
d^3{\vec k}_1 d^3{\vec k}_2 d^3{\vec k}_3}} 
\frac{\delta^3(\vec k_1+\vec k_2+\vec k_3)}{2\omega_1 2\omega_2 2\omega_3} \nonumber \\
&&
\times\Psi^\star_{PJJ_3'}\left({\vec k}_1,{\vec k}_2,{\vec k}_3;
\sigma'_1,\sigma'_2,\sigma'_3\right) \nonumber \\
&&
\times\left<k_1,k_2,k_3;\sigma'_1,\sigma'_2,\sigma'_3\right|\hat{A}^{\mu}_3
\left|k_1,k_2,k_3;\sigma_1,\sigma_2,\sigma_3\right>
\nonumber\\
&&
\times\Psi_{PJJ_3}\left({\vec k}_1,{\vec k}_2,{\vec k}_3;
\sigma_1,\sigma_2,\sigma_3\right)  \, .
\label{transampl}
\end{eqnarray}
The $\Psi$'s are the momentum-space representations of the baryon eigenstates for
$\vec P=0$, i.e. the rest-frame wave functions of the baryon ground and resonance states with corresponding mass $M$ and total angular momentum $J$ and $z$-projections $J_3$
as well as $J_3'$. Here they are expressed as functions of the individual quark
three-momenta $\vec k_i$, which sum up to $\vec P=\vec k_1+\vec k_2+\vec k_3=0$;
$\omega_i=\sqrt{m^2_i+\vec k^2_i}$ is the energy of quark $i$ with mass $m_i$, and the
individual-quark spin orientations are denoted by $\sigma_i$.

The integral on the r.h.s. of Eq.~(\ref{transampl}) is evaluated along the SM what
amounts to the matrix element of the axial current operator
$\hat{A}^{\mu}_a$ between (free) three-particle states
$\left|k_1,k_2,k_3;\sigma_1,\sigma_2,\sigma_3\right>$  to be assumed in the form
\begin{multline}
\left<k_1,k_2,k_3;\sigma'_1,\sigma'_2,\sigma'_3\right|
{\hat A}^{\mu}_3
\left|k_1,k_2,k_3;\sigma_1,\sigma_2,\sigma_3\right>
= \\
3\left<k_1,\sigma'_1\right|\hat{A}^{\mu}_{3,{\rm SM}}
\left|k_1,\sigma_1\right>2\omega_2 2\omega_3
\delta_{\sigma_{2}\sigma'_{2}}\delta_{\sigma_{3}\sigma'_{3}}
\label{eq:axcurr1}\, .
\end{multline}
For point-like quarks this matrix element involves the axial current operator
of the active quark 1 (with quarks 2 and 3 being the spectators) in the form
\begin{equation}
%\begin{multline}
\left<k_1,\sigma'_1\right|\hat{A}^{\mu}_{3,{\rm SM}}
\left|k_1,\sigma_1\right>= 
%\\
{\bar u}\left(k_1,\sigma'_1\right)g_A^q \gamma^\mu
\gamma_5 \frac{{\tau}_3}{2} u\left(k_1,\sigma_1\right) \, ,
\label{eq:axcurr2}
%\end{multline}
\end{equation}
where $u\left(k_1,\sigma_1\right)$ is the spinor of a quark with flavor $u$ or $d$
and $g_A^q=1$ its axial charge. A pseudovector current analogous to the one in
Eq.~(\ref{eq:axcurr2}) was recently
also used in the calculation of $g_{\pi NN}$ and the strong $\pi NN$ vertex form factor
in ref.~\cite{Melde:2008dg}.

For the calculation of the axial charges $g_A$ we can use either one of the components
$\mu=i=1, 2, 3$ of the axial current operator $\hat{A}^{\mu}_{3,{\rm SM}}$ in
Eq.~(\ref{eq:axcurr2}). The expression on the r.h.s. then specifies to
\begin{eqnarray}
&&{\bar u}\left(k_1,\sigma'_1\right) \gamma^i\gamma_5 \frac{{\tau}_3}{2} u\left( k_1,\sigma_1\right)=\nonumber\\
&&
2\omega_1\chi^*_{\frac{1}{2},\sigma'_1}
\Biggl\{\left[1-\frac{2}{3}\left(1-\kappa\right) \right]\sigma^i \nonumber\\
&&
+\sqrt{\frac{5}{3}}\frac{\kappa^2}{1+\kappa}
\left[\,\left[\vec{v}_{1}\otimes\vec{v}_{1}\right]_2\otimes\vec{\sigma}\right]_1^i\Biggl\}
\frac{{\tau}_3}{2} \chi_{\frac{1}{2},\sigma_1}\,,
\label{eq:ga}
\end{eqnarray}
where $\kappa=1/\sqrt{1+v_1^2}$ and $\vec v_1=\vec k_1/m_1$.
Herein $\sigma^i$ is the $i$-th component of the usual Pauli matrix
$\vec \sigma$ and $v_1$ the magnitude of the three-velocity $\vec v_1$. The symbol
$\left[.\otimes .\right]_k^i$ denotes the $i$-th component of a tensor product
$\left[.\otimes .\right]_k$ of rank $k$. We note that a similar formula was already
published before by Dannbom et al.~\cite{Dannbom:1996sh}, however, restricted to the
case of total orbital angular momentum $L=0$. Our expression holds for any $L$, thus
allowing to calculate $g_A$ for the most general wave function of a baryon ground or
resonances state specified by intrinsic spin and parity $J^P$.

\section{Results}

\renewcommand{\arraystretch}{1.4}
\begin{table*}
\centering
\caption{Axial charges $g_A^B$ of octet and decuplet ground states as predicted by the
EGBE~\cite{Glantschnig:2004mu},
psGBE~\cite{Glozman:1997ag}, and OGE~\cite{Theussl:2000sj} RCQMs in comparison to
experiment~\cite{Amsler:2008zzb} and lattice QCD results from Lin and Orginos
(LO)~\cite{Lin:2007ap} and
Erkol, Oka, and Takahashi (EOT)~\cite{Erkol:2009ev} as well as $\chi$PT results from
Jiang and Tiburzi (JT)~\cite{Jiang:2008we,Jiang:2009sf}; also given is the
nonrelativistic limit (NR) from the EGBE RCQM.}
\begin{tabular}{ccccccccc}
\hline 
\hline
 & Exp & EGBE & psGBE & OGE  &LO&EOT&JT& NR\tabularnewline
\hline 
N & 1.2695$\pm$0.0029 & 1.15 & 1.15 & 1.11  & 1.18$\pm$0.10&1.314$\pm$0.024&1.18& 1.65\tabularnewline
$\Sigma$ & - & 0.65 & 0.65 & 0.65  & 0.636$\pm$0.068$^\dagger$
&0.686$\pm$0.021$^\dagger$&0.73& 0.93\tabularnewline
$\Xi$ & - & -0.21 & -0.22 & -0.22  & -0.277$\pm$0.034&-0.299$\pm$0.014$^\ddagger$&-0.23$^\ddagger$& -0.32\tabularnewline
\hline 
$\Delta$ & - & -4.48 & -4.47 & -4.30 &-&-&$\sim\,$-4.5& -6.00 \tabularnewline
$\Sigma^{*}$ & - & -1.06 & -1.06 & -1.00  & -&-&-& -1.41\tabularnewline
$\Xi^{*}$ & - & -0.75 & -0.75 & -0.70  & -&-&-& -1.00\tabularnewline
\hline
\hline
\end{tabular}
\\
\begin{flushleft}
$^\dagger$ Due to another definition of $g_A^{\Sigma}$ this numerical value is
different by a $\sqrt{2}$ from the one quoted in the original paper.\\
$^\ddagger$ Due to another definition of $g_A^{\Xi}$ this value has a sign opposite
to the one in the original paper.
\end{flushleft}
\label{EGBE}
\end{table*}

In Table~\ref{EGBE} we present the RCQM results from our calculations for the axial
charges $g_A^B$ of the octet and decuplet ground states
$B=N$, $\Sigma$, $\Xi$, $\Delta$, $\Sigma^*$, and $\Xi^*$. Except for the $N$
there are no direct
experimental data for $g_A$ (from $\Delta S=0$ decays). The predictions for $g_A^N$
by all three RCQMs come close to the experimental value, with all of them falling slightly
below it. This is also the trend of present-day lattice-QCD calculations~\cite{Renner_2009};
only the EOT result seems to represent a notable exception, even if we take the
theoretical uncertainties into account (in Table~\ref{EGBE} we have chosen to quote the
EOT result corresponding to their calculation with the smallest quark mass of 35 MeV).
In addition, also the
JT prediction obtained from $\chi$PT remains below the experimental value. There might be
a variety of reasons why the different approaches underestimate the $g_A^N$. However,
one should also bear in mind that the phenomenological value of $g_A^N\sim\,$1.27 is
supposed under the conjecture of conserved vector currents. What concerns the RCQMs,
and in particular the {\mbox psGBE} RCQM, interestingly, it has recently been
found~\cite{Melde:2004qu}
that also the $\pi NN$ coupling constant turns out to be too small, namely
$\frac{f^2_{\pi NN}}{4\pi}=$ 0.0691, as compared to the phenomenological value of
about 0.075~\cite{Bugg:2004cm}. It remains to be clarified if in case of the
RCQMs these undershootings of both the $g_A^N$ and $f^2_{\pi NN}$, which are related
by the Goldberger-Treiman relation, have to be interpreted as lacking $\pi-$dressing
effects.

In the last column of Table~\ref{EGBE} we quote also the nonrelativistic limit of the
prediction by the EGBE RCQM (i.e. for the limit $\kappa \rightarrow$ 1 in Eq. (\ref{eq:ga})).
It deviates grossly from the relativistic result, indicating that a consideration of
axial charges within a nonrelativistic approach is unreliable. This conclusion is
further substantiated by considering the axial charges of $N$ resonances, for which
indeed no experimental data are available but lattice-QCD results have recently been
produced. While the relativistic predictions of especially the EGBE RCQM agree very
well with the lattice-QCD data in case of both the $N$(1535) and $N$(1650) resonances,
the nonrelativistic limits deviate here too~\cite{Choi2010}.

For the $g_A^B$ of the octet and decuplet ground states the RCQMs yield very similar
results. While the predictions of the psGBE and the EGBE are essentially the same,
differences occur only for the OGE RCQM, but they remain within at most $\sim\,$6\%.
In case of the octet states $\Sigma$ and $\Xi$ we can also compare to lattice-QCD
as well as $\chi$PT results. The comparison of the RCQM predictions to the former
is quite satisfying, as the figures agree rather well. Except for $g_A^{\Sigma}$
practically the same is true with regard to the $\chi$PT results of JT. Again,
the results from the nonrelativistic limit of the EGBE RCQM fall short; as in
the case of the $N$ the corresponding values are always bigger (in absolute value)
than all of the other results.

For the decuplet ground states $\Delta$, $\Sigma^*$, and $\Xi^*$
there are neither experimental data nor lattice-QCD
results. Only for the $\Delta$ we may compare with a $\chi$PT prediction, showing
again a striking similarity. For the other cases of $\Sigma^*$ and $\Xi^*$ we have
here produced first predictions and one has still to await results from other approaches.

\begin{table}[h]
\centering
\caption{Mass eigenvalues and axial charges $g_A^N$ of the $N$ ground state and the
low-lying $N$ resonances as predicted by the EGBE, the psGBE, and the OGE RCQMs.}
\begin{tabular}{lcp{3mm}crp{3mm}crp{3mm}cr}
\hline\hline \\[-3mm]
\multicolumn{2}{c}{} && \multicolumn{2}{c}{EGBE} && \multicolumn{2}{c}{psGBE} && \multicolumn{2}{c}{OGE}\tabularnewline
\hline 
State & $J^{p}$ && Mass & $g_{A}$ && Mass & $g_{A}$ && Mass & $g_{A}$\tabularnewline
\hline\\[-2mm]  
$N$(939) & $\frac{1}{2}^{+}$ && 939 & 1.15 && 939& 1.15 && 939 & 1.11\tabularnewline

$N$(1440) & $\frac{1}{2}^{+}$ && 1464 & 1.16 && 1459 & 1.13 && 1578 & 1.10\tabularnewline

$N$(1520) & $\frac{3}{2}^{-}$ && 1524 & -0.64 && 1519 & -0.21 && 1520 & -0.15\tabularnewline

$N$(1535) & $\frac{1}{2}^{-}$ && 1498 & 0.02 && 1519 & 0.09 && 1520 & 0.13\tabularnewline

$N$(1650) & $\frac{1}{2}^{-}$ && 1581 & 0.51 && 1647 & 0.46 && 1690 & 0.44\tabularnewline

$N$(1700) & $\frac{3}{2}^{-}$ && 1608 & -0.10 && 1647 & -0.50 && 1690 & -0.47\tabularnewline

$N$(1710) & $\frac{1}{2}^{+}$ && 1757 & 0.35 && 1776 & 0.37 && 1860 & 0.32\tabularnewline

$N$(1720) & $\frac{3}{2}^{+}$ && 1746 & 0.35 && 1728 & 0.34 && 1858 & 0.25\tabularnewline

$N$(1675) & $\frac{5}{2}^{-}$ && 1676 & 0.84 && 1647 & 0.83 && 1690 & 0.80\tabularnewline

$N$(1680) & $\frac{5}{2}^{+}$ && 1689 & 0.89 && 1728& 0.83 && 1858 & 0.70\tabularnewline
\hline
\hline
\end{tabular}
\label{N}
\end{table}

Next we come to discuss the axial charges of nucleon and other baryon resonances.
As mentioned in the Introduction, especially the nucleon resonances have recently
attracted interest, mainly because the issue of chiral-symmetry restoration
higher in the baryon spectra has been raised~\cite{Glozman:2007ek,Glozman:2008vg}
and because first lattice-QCD calculations have been performed~\cite{Takahashi:2008fy}.
Certainly the results of the latter have still to be taken with care, as they
correspond to relatively high quark masses. For the case of the nucleon we have
presented resonance axial charges from RCQMs already
in a previous paper~\cite{Choi2010}; for completeness we repeat them here in
Table~\ref{N}. While for details we refer to ref.~\cite{Choi2010}, we mention
as the main characterization of these results that \\[2mm]
$i$) the RCQM predictions perfectly agree with the lattice-QCD results
for the $N$(1535) and $N$(1650) resonances, i.e. in the two cases for which
lattice-QCD calculations have so far become available, \\[2mm]
$ii$) the small, practically vanishing, $g_A$ of $N$(1535) can be reproduced with
$\{QQQ\}$ configurations alone, \\[2mm]
$iii$) the predictions of different RCQMs are generally very similar except for the
$J^P=\frac{3}{2}^-$ resonances $N$(1520) and $N$(1700), \\[2mm]
$iv$) a relativistic description is necessary and a simple $SU(6)\times O(3)$
nonrelativistic quark model is not reliable, and \\[2mm]
$v$) there is no tendency of the axial charges of high-lying parity partners to
assume particularly small values.

\begin{table}[h]
\centering
\caption{Same as Table~\ref{N} but for the octet $\Sigma$ and decuplet $\Sigma^*$ states.}
\begin{tabular}{lcp{3mm}crp{3mm}crp{3mm}cr}
\hline\hline \\[-3mm]
\multicolumn{2}{c}{} && \multicolumn{2}{c}{EGBE} && \multicolumn{2}{c}{psGBE} && \multicolumn{2}{c}{OGE}\tabularnewline
\hline 
State & $J^{p}$ && Mass & $g_{A}$ && Mass & $g_{A}$ && Mass & $g_{A}$\tabularnewline
\hline\\[-2mm]  
$\Sigma$(1193) & $\frac{1}{2}^{+}$ && 1194 & 0.65 && 1182 & 0.65 && 1121 & 0.65\tabularnewline

$\Sigma$(1560) & $\frac{1}{2}^{-}$ && 1672  & -0.15 && 1678& -0.07&& 1655 & 0.01\tabularnewline

$\Sigma$(1620) & $\frac{1}{2}^{-}$ && 1740  & 0.62 && 1736& 0.58&& 1770 & 0.54\tabularnewline

$\Sigma$(1660) & $\frac{1}{2}^{+}$ &&  1664 & 0.69 && 1619 & 0.64 && 1755 & 0.64 \tabularnewline

$\Sigma$(1670) & $\frac{3}{2}^{-}$ && 1681  & -0.92 && 1678& -0.48 && 1655 & -0.24 \tabularnewline

$\Sigma$(1775) & $\frac{5}{2}^{-}$ && 1765  & 1.06 && 1736&  1.03&& 1770 & 0.97 \tabularnewline

$\Sigma$(1880) & $\frac{1}{2}^{+}$ && 1903  & 0.38 && 1912 & 0.42  && 1980 & 0.17\tabularnewline

$\Sigma$(1940) & $\frac{3}{2}^{-}$ && 1725  & -0.45 && 1736& -0.83&& 1770 & -0.78 \tabularnewline

\hline

$\Sigma^*$(1385) & $\frac{3}{2}^{+}$ && 1365  & -1.06 && 1389&  -1.06&& 1311 & -1.00\tabularnewline

$\Sigma^*$(1690) & $\frac{3}{2}^{+}$ && 1812  & -1.05 && 1865&  -1.03&& 1932 & -0.99 \tabularnewline

$\Sigma^*$(1750) & $\frac{1}{2}^{-}$ && 1761  & -0.08  && 1759 & -0.13 && 1718 & -0.18\tabularnewline
\hline
\hline
\end{tabular}
\label{Sigma}
\end{table}

The RCQM predictions for the axial charges of the octet $\Sigma$ and decuplet
$\Sigma^*$ resonances are quoted in Table~\ref{Sigma}. The gross pattern of the
results is like the one of the $N$ resonances. First of all the different
{\mbox RCQMs} yield similar values for the axial charges except for the cases of
$\Sigma$(1670) and $\Sigma$(1940), which are again $J^P=\frac{3}{2}^-$ resonances
and are to be assigned to the same octets as $N$(1520) and $N$(1700), respectively,
according to a recent identification of baryon resonances~\cite{Melde:2008yr}
(see also~\cite{PDG2010}). For both $\Sigma$(1670) and $\Sigma$(1940) the differences
among the predictions prevail also in the comparison between the EGBE and the
psGBE RCQMs hinting to considerable influences from tensor and/or spin-orbit forces,
just as in the corresponding two $N$ resonances. All of the RCQMs
produce very small values for the axial charges of $\Sigma$(1560). Notably,
this state falls into the same octet as $N$(1535)~\cite{Melde:2008yr,PDG2010},
whose $g_A$ is also extremely small (see Table~\ref{N}). A similar small axial
charge is found for the decuplet $\Sigma^*$(1750). In general, however, we do not
observe the axial charges to become small as we go up to higher resonances.

The results for the axial charges of the octet $\Xi$ and decuplet $\Xi^*$ resonances
are collected in Table~\ref{Xi}. Here, all the RCQMs produce similar predictions,
where the axial charges of the octet resonances remain rather small with values
ranging from -0.2 to -0.4.

\begin{table}[h]
\centering
\caption{Same as Table~\ref{N} but for the octet $\Xi$ and decuplet $\Xi^*$ states..}
\begin{tabular}{lcp{3mm}crp{3mm}crp{3mm}cr}
\hline\hline \\[-3mm]
\multicolumn{2}{c}{} && \multicolumn{2}{c}{EGBE} && \multicolumn{2}{c}{psGBE} && \multicolumn{2}{c}{OGE}\tabularnewline
\hline 
State & $J^{p}$ && Mass & $g_{A}$ && Mass & $g_{A}$ && Mass & $g_{A}$\tabularnewline
\hline\\[-2mm]  
$\Xi$(1318) & $\frac{1}{2}^{+}$ && 1355  & -0.21 && 1348 & -0.22 && 1193 & -0.22\tabularnewline

$\Xi$(1690) & $\frac{1}{2}^{+}$ && 1813  & -0.23 && 1806 & -0.22 && 1826 & -0.22\tabularnewline

$\Xi$(1820) & $\frac{3}{2}^{-}$ && 1807  & -0.38 && 1792 & -0.40 && 1751 & -0.23 \tabularnewline

\hline

$\Xi^*$(1530) & $\frac{3}{2}^{+}$ && 1512  & -0.75 && 1528 & -0.75  && 1392 & -0.70 \tabularnewline
\hline
\hline
\end{tabular}
\label{Xi}
\end{table}

Finally, in Table~\ref{Delta} the axial charges of the $\Delta$ resonances are given.
Again, all of the RCQMs yield similar predictions. Only it is remarkable that the
axial charges especially of the $J^P=\frac{3}{2}^+$ states are rather big in
absolute value. If we consider the $\Delta$(1232) ground state, its $g_A$ is at
least three times larger than the one of the $N$ ground state. Remarkably a ratio
of about the same size has recently been found between the $\pi N\Delta$ and the
$\pi NN$ strong coupling constants~\cite{Melde:2008dg}. The smallest $g_A$ is
found for $\Delta$(1620). It should be noted that it falls into the same decuplet
as the $\Sigma^*$(1750), whose $g_A$ was also seen as the smallest among the
$\Sigma^*$ resonances (cf. Table~\ref{Sigma}).

\begin{table}[h]
\centering
\caption{Same as Table~\ref{N} but for the $\Delta$ states.}
\begin{tabular}{lcp{3mm}crp{3mm}crp{3mm}cr}
\hline\hline \\[-3mm]
\multicolumn{2}{c}{} && \multicolumn{2}{c}{EGBE} && \multicolumn{2}{c}{psGBE} && \multicolumn{2}{c}{OGE}\tabularnewline
\hline 
State & $J^{p}$ && Mass & $g_{A}$ && Mass & $g_{A}$ && Mass & $g_{A}$\tabularnewline
\hline\\[-2mm]  
$\Delta$(1232) & $\frac{3}{2}^{+}$ && 1231  & -4.48 && 1240 & -4.47 && 1231 & -4.30\tabularnewline

$\Delta$(1600) & $\frac{3}{2}^{+}$ && 1686  & -4.41 && 1718 & -4.33 && 1855 & -4.20 \tabularnewline

$\Delta$(1620) & $\frac{1}{2}^{-}$ && 1640  & -0.76 && 1642 & -0.75 && 1621 & -0.74\tabularnewline

$\Delta$(1700) & $\frac{3}{2}^{-}$ && 1639  & -1.68 && 1642 & -1.66 && 1621 & -1.58\tabularnewline

\hline
\hline
\end{tabular}
\label{Delta}
\end{table}

\section{Conclusions}

We have presented results from a comprehensive and consistent study of axial charges
$g_A^B$ of octet and decuplet baryon ground and resonant states with RCQMs. The
dynamical models differ mainly with regard to their hyperfine $Q$-$Q$ interactions,
which stem from OGE and psGBE as well as EGBE. Whenever a comparison is possible
with either experimental data or established theoretical results (especially from
lattice QCD and $\chi$PT), the RCQM predictions turn out to be quite reasonable.
The values deduced from a nonrelativistic approximation in general differ grossly,
indicating that a relativistic approach to the axial charges is mandatory. The RCQMs
considered here rely on $\{QQQ\}$ configurations only. Nevertheless, the $g_A^B$
results never fall short but rather produce a consistent picture. Already for the
ground states one finds a scatter of $g_A^B$ values from small to large. Through
the Goldberger-Treiman relation one may thus expect smaller or larger $\pi$-dressing
effects depending on the baryon state. Particularly big are the axial charges of the
$\delta$ ground and first excited states, much in congruency with the relatively
large $\pi N\Delta$ coupling constant. The axial charges of some baryon resonances
are rather sensitive to tensor and/or spin-orbit forces in the hyperfine interaction.
These resonances fall into the same flavor multiplets. In general the pattern observed
from the predictions for $g_A^B$ is congruent with the classification of baryons
into flavor multiplets as found recently. From the RCQMs predictions presented
here, no particular trend is observed for the axial charges of $N$ and other
baryon resonances to become small, when the excitation energy is increased.
Certainly, the consideration of baryon axial charges remains an exciting field,
and one is eager to see additional experimental data as well as more theoretical
results from different approaches to QCD.

\begin{acknowledgments}
The authors are grateful to L.Ya. Glozman for valuable incentives regarding specific
aspects of this work and to the Graz lattice-QCD group for several clarifying discussions
about respective calculations.
This work was supported by the Austrian Science Fund, FWF, through the Doctoral
Program on {\it Hadrons in Vacuum, Nuclei, and Stars} (FWF DK W1203-N08).
\end{acknowledgments}

\appendix*

\section{Rarita-Schwinger spinors}

For the $J=\frac{1}{2}$, $\frac{3}{2}$, and $\frac{5}{2}$ baryons with four-momentum
$P$ and energy $E=\sqrt{M^2+{\vec P}^2}$ we employ Dirac
and Rarita-Schwinger spinors, similar as in ref.~\cite{Choi:2007gy}, as follows:
\begin{widetext}
$\bullet \;\; J=\frac{1}{2}, \;\; U\left(P,J_3=\pm\frac{1}{2}\right)\,:$
\begin{eqnarray}
U\left(P,\frac{1}{2}\right)=\left(\begin{array}{c}
\sqrt{E+M}\\0\\ \frac{{\vec\sigma} \cdot{\vec P}}{\sqrt{E+M}}\\0\end{array}\right) ,
\hspace{7mm} U\left(P,-\frac{1}{2}\right)=\left(\begin{array}{c}
0\\\sqrt{E+M}\\ 0\\ \frac{{\vec\sigma} \cdot{\vec P}}{\sqrt{E+M}}\end{array}\right) ,
\end{eqnarray}
where $\vec\sigma$ are the Pauli matrices. These Dirac spinors are normalized as 
\begin{equation}
\bar{U}(P,J_3') U(P,J_3)=\delta_{J_3', J_3}2M\, .
\end{equation}
$\bullet \;\; J=\frac{3}{2}, \;\;
U^\mu \left(P,J_3=\pm\frac{1}{2},\pm\frac{3}{2}\right)\,:$
\begin{eqnarray}
U^{\mu}\left(P,\frac{3}{2}\right)&=&e^{\mu}_{+}(P)U\left(P,\frac{1}{2}\right)\, ,
\nonumber\\
U^{\mu}\left(P,\frac{1}{2}\right)&=&\sqrt{\frac{2}{3}}e^{\mu}_{0}(P)U\left(P,\frac{1}{2}\right) 
+\sqrt{\frac{1}{3}}e^{\mu}_{+}(P)U\left(P,-\frac{1}{2}\right)\, ,
\nonumber\\
U^{\mu}\left(P,-\frac{1}{2}\right)&=&\sqrt{\frac{1}{3}}e^{\mu}_{-}(P)U\left(P,\frac{1}{2}\right)
+\sqrt{\frac{2}{3}}e^{\mu}_{0}(P)U\left(P,-\frac{1}{2}\right)\, ,
\nonumber\\
U^{\mu}\left(P,-\frac{3}{2}\right)&=&e^{\mu}_{-}(P)U\left(P,-\frac{1}{2}\right)\, .
\end{eqnarray}
$\bullet \;\; J=\frac{5}{2}, \;\;
U^{\mu\nu}\left(P,J_3=\pm\frac{1}{2},\pm\frac{3}{2},\pm\frac{5}{2}\right)\,:$
\begin{eqnarray}
U^{\mu\nu}\left(P,\frac{5}{2}\right)&=&e^{\mu}_{+}e^{\nu}_{+}U\left(P,\frac{1}{2}\right)\, ,
\nonumber\\
U^{\mu\nu}\left(P,\frac{3}{2}\right)&=&\sqrt{\frac{2}{5}}e^{\mu}_{+}
e^{\nu}_{0}U\left(P,\frac{1}{2}\right)+\sqrt{\frac{1}{5}}e^{\mu}_{+}e^{\nu}_{+}
U\left(P,-\frac{1}{2}\right)
+\sqrt{\frac{2}{5}}e^{\mu}_{0}e^{\nu}_{+}U\left(P,\frac{1}{2}\right)\, ,\nonumber\\
U^{\mu\nu}\left(P,\frac{1}{2}\right)&=&\sqrt{\frac{1}{10}}e^{\mu}_{+}
e^{\nu}_{-}U\left(P,\frac{1}{2}\right)+\sqrt{\frac{1}{5}}e^{\mu}_{+}e^{\nu}_{0}
U\left(P,-\frac{1}{2}\right)
+\sqrt{\frac{2}{5}}e^{\mu}_{0}e^{\nu}_{0}U\left(P,\frac{1}{2}\right)\nonumber\\
&&+\sqrt{\frac{1}{5}}e^{\mu}_{0}e^{\nu}_{+}U\left(P,-\frac{1}{2}\right)
+\sqrt{\frac{1}{10}}e^{\mu}_{-}e^{\nu}_{+}U\left(P,\frac{1}{2}\right)\, ,
\nonumber\\
U^{\mu\nu}\left(P,-\frac{1}{2}\right)&=&\sqrt{\frac{1}{10}}e^{\mu}_{+}
e^{\nu}_{-}U\left(P,-\frac{1}{2}\right)+\sqrt{\frac{1}{5}}e^{\mu}_{0}e^{\nu}_{-}
U\left(P,\frac{1}{2}\right)
+\sqrt{\frac{2}{5}}e^{\mu}_{0}e^{\nu}_{0}U\left(P,-\frac{1}{2}\right)\nonumber \\
&&+\sqrt{\frac{1}{5}}e^{\mu}_{-}e^{\nu}_{0}
U\left(P,\frac{1}{2}\right)
+\sqrt{\frac{1}{10}}e^{\mu}_{-}e^{\nu}_{+}
U\left(P,-\frac{1}{2}\right)\, ,\nonumber\\
U^{\mu\nu}\left(P,-\frac{3}{2}\right)&=&\sqrt{\frac{2}{5}}e^{\mu}_{0}
e^{\nu}_{-}U\left(P,-\frac{1}{2}\right)+\sqrt{\frac{1}{5}}e^{\mu}_{-}
e^{\nu}_{-}U\left(P,-\frac{1}{2}\right)
+\sqrt{\frac{2}{5}}e^{\mu}_{-}e^{\nu}_{0}U\left(P,-\frac{1}{2}\right)\, ,
\nonumber\\
U^{\mu\nu}\left(P,-\frac{5}{2}\right)&=&e^{\mu}_{-}e^{\nu}_{-}
U\left(P,-\frac{1}{2}\right)\, .
\end{eqnarray}
In the latter equation we have suppressed the arguments in the polarization
vectors $e^{\mu}_{\lambda}(P)$ defined by
\begin{equation}                
e^{\mu}_{\lambda}(P)=\left(\frac{\hat{e}_{\lambda}\cdot
{\vec P}}{M},\, \hat{e}_{\lambda}
+\frac{(\hat{e}_{\lambda}\cdot{\vec P}){\vec P}}{M(E+M)}\right),
\end{equation}
where for $\lambda=+, 0, -$ the unit vectors $\hat e_\lambda$ are written as
\begin{equation}
\hat{e}_{+}=-\frac{1}{\sqrt{2}}(1,i,0),\,\hat{e}_{0}=(0,0,1),\,
\hat{e}_{-}
=\frac{1}{\sqrt{2}}(1,-i,0)\, .
\end{equation}  
\end{widetext}

\addcontentsline{toc}{chapter}{Bibliography}
\bibliographystyle{prsty}

\end{document}